\documentclass[twocolumn,aps,prl,showpacs,superscriptaddress]{revtex4}
\usepackage{graphicx}
\usepackage{amssymb} 
\usepackage{url}     
\usepackage{bm}      
\usepackage{color}
\usepackage{float}
\usepackage{natbib}

\newcommand{\bise}{Bi$_2$Se$_3$}

\newcommand{\didv}{$dI/dV$}
\newcommand{\dvdi}{$dV/dI$}

\newcommand{\sio}{SiO$_2$}

\begin{document}

\author{A. Zalic}
\affiliation{Racah Institute of Physics, The Hebrew University, Jerusalem 91904, Israel}
\author{T. Dvir}
\affiliation{Racah Institute of Physics, The Hebrew University, Jerusalem 91904, Israel}
\author{H. Steinberg}
\affiliation{Racah Institute of Physics, The Hebrew University, Jerusalem 91904, Israel}


\title{High density carriers at a strongly coupled graphene-topological insulator interface}

\begin{abstract}
We report on a strongly coupled bilayer graphene (BLG) - \bise\ device with a junction resistance of less than 1.5 k$\Omega\mu$m$^2$. This device exhibits unique behavior at the interface, which cannot be attributed to either material in absence of the other. We observe quantum oscillations in the magnetoresistance of the junction, indicating the presence of well-resolved Landau levels due to hole carriers of unknown origin with a very large Fermi surface. These carriers, found only at the interface, could conceivably arise due to significant hole doping of the bilayer graphene with charge transfer on the order of 2$\times$10$^{13}$ cm$^{-2}$, or due to twist angle dependent mini-band transport. 
\end{abstract}

\maketitle

\date{\today}

The interface between graphene and the surface of a strong three dimensional topological insulator (TI) is an attractive model system for probing proximity effects involving topological states. Studies predict the possibility of strong band hybridization, where graphene attains an enhanced spin-orbit coupling \cite{Zhang2014b,Jin2013a} and non-trivial spin-textured bands \cite{Zhang2014b}. Band hybridization may lead to the formation of a heavy Dirac fermion band \cite{Cao2016}, the appearance of a topologically protected hybrid state \cite{Popov2014}, or the emergence of topological order in the graphene \cite{Kou2013}. At the same time, other studies suggest the interface is weakly coupled \cite{Liu2013a,Spataru2014a}, with finite charge transfer between the two surfaces. In this case graphene forms a non-disruptive contact to the TI, which retains its spin-momentum locking. This can then be used for spin-injection into graphene \cite{Li2012} or mutual tunneling measurements \cite{Steinberg2015}.

Thus far, the interface between graphene and the topological insulator \bise\ has been shown by experiment to be a tunneling interface, with the differential conductance reflecting both graphene and \bise\ density of states (DOS) \cite{Steinberg2015,Zhang2016a}. Recent experimental studies report use of the topological insulator Bi$_2$Te$_2$Se (BTS) to induce spin current in graphene with one of these reporting an ohmic interface between epitaxially grown BTS on graphene, hinting at strong coupling  \cite{Vaklinova2016a,tian2016electrical}. However, no clear experimental signature has been reported which indicates an alteration in properties of either graphene or TI when they are in proximity to each other. Thus, whether or not strong coupling and hybridization may be achieved in a graphene-TI heterostructure remains an open question.

\renewcommand\floatpagefraction{.1}
\begin{figure}[h!]
	
	\begin{center}
		\includegraphics[width = 80mm]{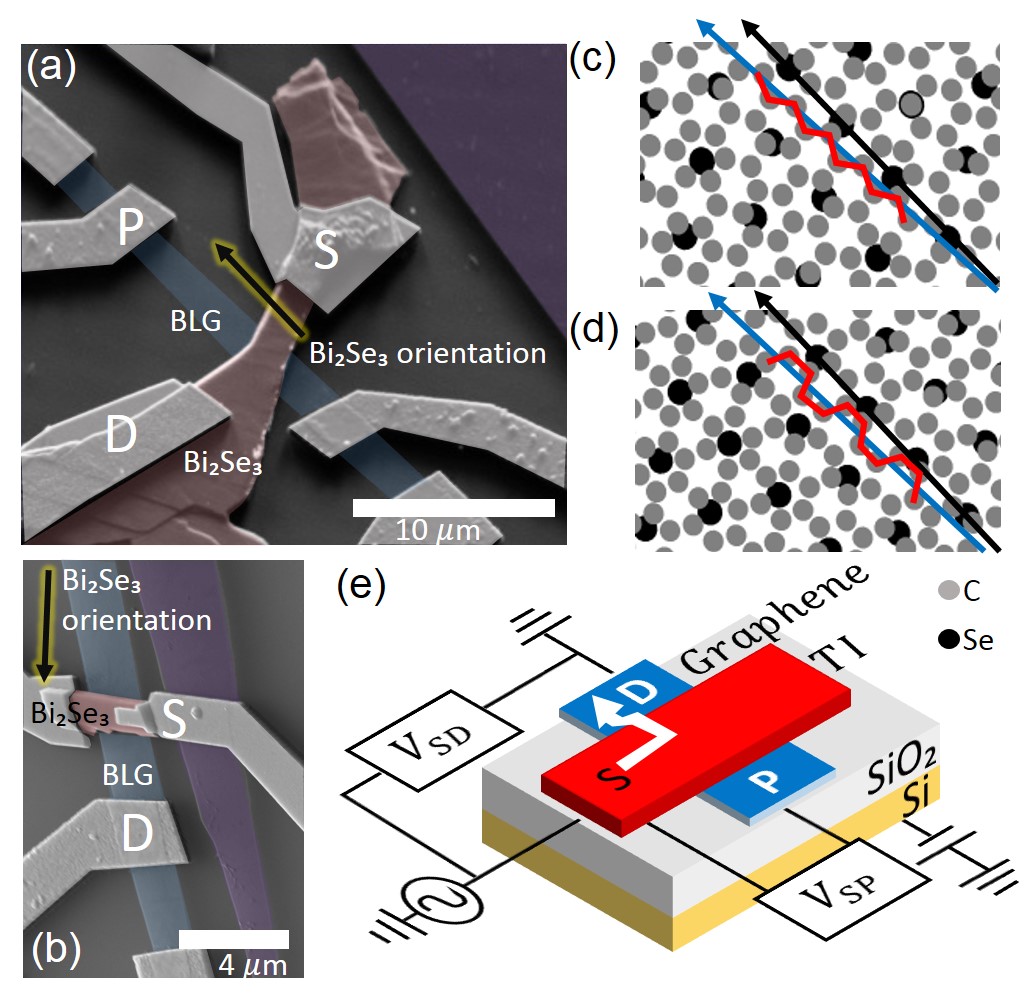}
		\vspace{0cm}
		\caption{(a),(b) False color SEM images of Devices 1,2, with BLG flake (blue) on the bottom and \bise\ flake (red) on top. Black arrow marks crystallographic orientation of \bise\ determined by EBSD measurement. (c),(d) Zigzag, armchair orientations between the BLG lattice (gray, top layer only shown) and \bise\ (black, bottom Se layer only shown) in Device 1. The ``airmchair'' orientation is at a nearly commensurate angle of 33$^\circ$. (e) Schematic illustration of measurement configuration. Current is induced between the source electrode on the TI and the drain electrode on the BLG, and voltage is measured between the source and the drain ($V_{SD}$), and between the source and the probe ($V_{SP}$). Gate voltage is applied between the Si back gate and the device.}
		\label{Figure_1}
	\end{center}
	\vspace{-0.8 cm}
\end{figure}

The graphene-TI interface belongs to the wider family of van-der-Waals heterostructures. Of these, graphene-hexagonal boron nitride (hBN) and twisted bilayer graphene (BLG) heterostructures have been extensively studied. The first system is an interface between graphene and an insulator; the second, between two sheets of graphene which are rotated with respect to each other. The behavior of these systems is strongly dependent on the relative crystallographic orientation between neighboring lattices (which we call the twist angle). At small twist angles, close commensuration between the lattices produces a long wavelength moir\'{e} pattern. In the case of graphene on the insulating hBN, the moir\'{e} pattern creates a periodic super-lattice (SL) potential affecting the graphene. This leads to the formation of graphene mini-bands within the super-lattice Brillouin zone (SLBZ). The bands are separated by gaps at the SLBZ boundaries, which close at chirality protected momenta to form secondary Dirac points \cite{Park2008,Yankowitz2012a,Hunt2013b,Ponomarenko2013a}. In twisted BLG, this picture is made richer by interlayer hybridization \cite{LopesDosSantos2007a,Mele2010,Bistritzer2011c}. Recent ARPES and transport measurements of twisted BLG reveal a band structure consisting of two rotated graphene dispersions which fold into the SLBZ, and may hybridize in the vicinity of their crossings in momentum space \cite{Ohta2012,Schmidt2014a,Cao2016a,Kim2016}.  

The graphene-\bise\ interface, like twisted BLG, involves two Dirac conductors with lattices which can be arranged in a close to commensurate fashion. For these lattices, the nearest commensuration is achieved at a 30$^\circ$ twist angle, as the \bise\ lattice constant of 4.138 \r{A} is equal to $\sqrt{3}$ times the graphene lattice constant of 2.46 \r{A}, with a mismatch of -3\%. Reference \cite{Song2010a} has indeed observed a moir\'{e} pattern of wavelength  7.1 nm in STM measurements of \bise\ grown epitaxially on a BLG substrate. Twist angle dependent behavior could introduce a possibility for both the formation of mini-bands and the emergence of strong coupling and hybridization around energies where the graphene and \bise\ bands intersect in momentum space. This provides strong motivation to carry out transport measurements on small twist angle devices. 

In this work we report transport measurements on a very low impedance BLG-\bise\ junction. At high magnetic fields, we observe quantum oscillations due to high density hole carriers at the interface which we cannot associate with either BLG or \bise\ alone. The combination of uncharacteristically low junction impedance with the presence of unusual charge carriers at the interface indicates strong coupling between the BLG and \bise. We use electron backscatter diffraction (EBSD) to determine the relative twist angle of the device, obtaining an estimate of around 3$^\circ$ modulo 30$^\circ$. We speculate that the observed strong coupling in our device may be due to fortuitously obtaining close commensuration between the two materials. The non-typical charge carriers at the interface might then stem from unusually large hole doping of the BLG, or from some form of super-lattice modulated transport.      

We present data from two BLG-\bise\ devices, which consist of a flake of bilayer graphene exfoliated on \sio\ with a $\sim$100 nm thick flake of \bise\ transferred over it in a cross configuration (Figs.~\ref{Figure_1}(a),(b)). The BLG was contacted and cleaned by heat annealing before mechanical transfer of the \bise. The \bise\ was contacted in a subsequent lithography step using ion-milling to remove the surface oxide. Further details of the fabrication procedure are given in the supplementary material~\cite{SI}. 

\begin{figure}[t]
	
	\begin{center}
		\includegraphics[width = 85mm]{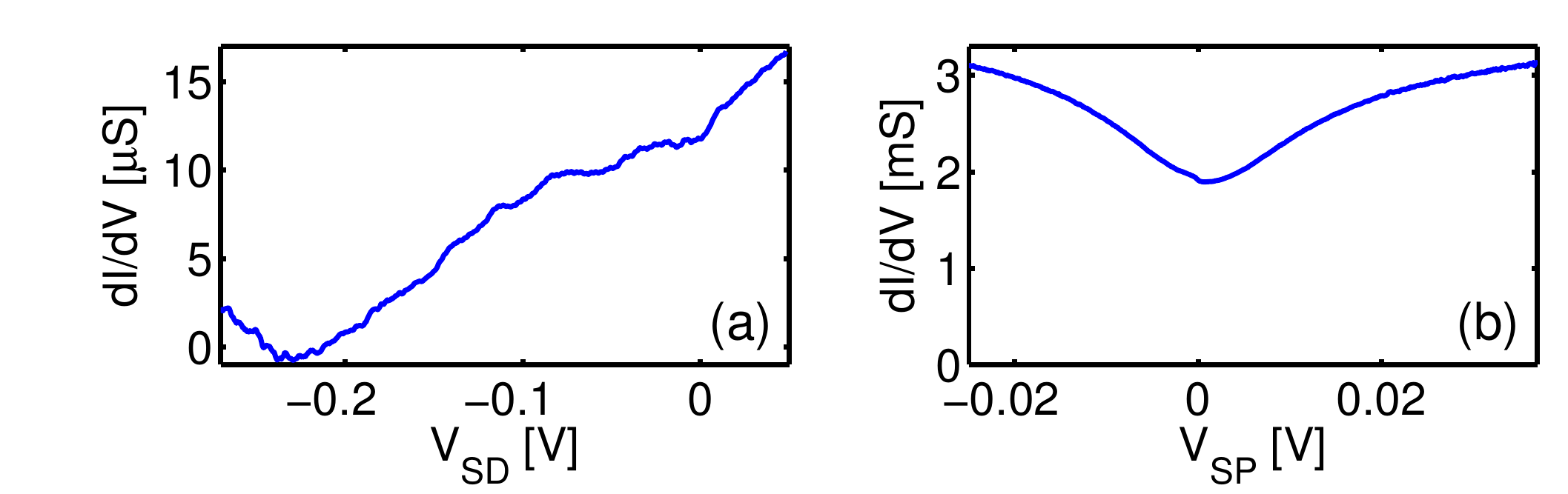}
		\vspace{0cm}
		\caption{ (a) $dI/dV$ vs. $V_{SD}$ of Device 2 (weakly coupled), tracing the linear DOS of the \bise\ surface state. The Dirac point, where the DOS reaches a minimum, is located at $V_{SD}\approx -230 mV$. (b). $dI/dV$ vs. $V_{SP}$ of Device 1 (strongly coupled). The trace is over a narrow range of $V_{SP}$ and does not exhibit any notable feature other than a low bias conductance suppression.   }
		\label{Figure_2}
	\end{center}
	\vspace{-0.3 cm}
\end{figure}

Devices 1 and 2, both fabricated on the same chip, exhibit very different interface resistances: Device 2 exhibits a typical tunneling resistance of $\sim$100 k$\Omega$ (250 k$\Omega\mu$m$^2$), whereas Device 1 shows an unusually low value of no more than 300 $\Omega$ (1.5 k$\Omega\mu$m$^2$). We study the interface by measuring the differential conductance (\didv ) or the differential resistance (\dvdi ) as a function of the bias voltage between the two materials, the gate voltage and the external magnetic field. Measurements were carried out at a temperature of $\approx$ 4.2 K. 

\begin{figure*}[t]
	
	\begin{center}
		\includegraphics[width = 176mm]{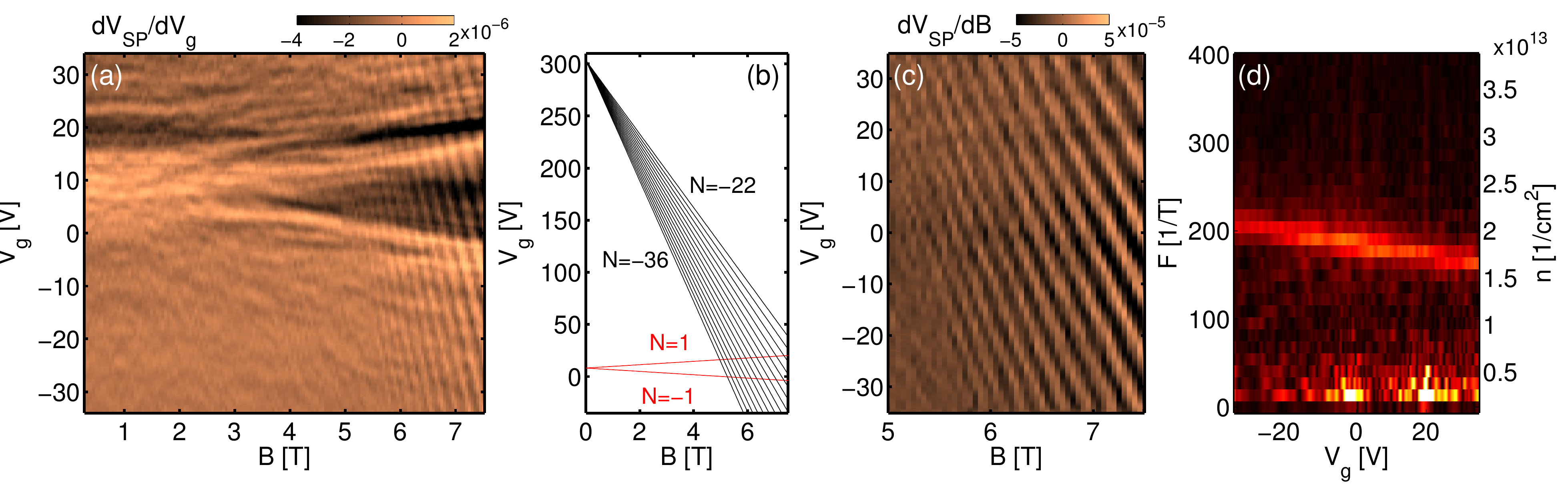}
		\vspace{0cm}
		\caption{Quantum oscillations at the junction of Device 1 in a perpendicular magnetic field. (a) $dV_{SP}/dV_g$ (color scale) vs. $V_g$ and $B$. Dispersing features may be traced to two intersecting Landau fans, originating at gate voltages of $\sim$8 V and $\sim$294 V. (b) Fits to the data in panel (a), extrapolated to the fan origins.  (c) $dV_{SP}/dB$ vs. $V_g$ and $B$, highlighting the fast oscillating Landau fan originating at $\sim$294 V. (d) Fourier amplitudes of $V_{SP}$ (linear background subtracted) vs. $V_g$ and frequency $F$ in $1/B$. $F$ is proportional to the area of the Fermi surface in momentum space $A_k$, which may be translated to the hole charge carrier density $n$ (right-hand y axis) assuming known Landau level degeneracy (here we take $g$ = 4 for BLG, see text).}
		\label{Figure_3}
	\end{center}
	\vspace{-0.5 cm}
\end{figure*}

Device 2, with its high junction impedance, was measured in a simple two terminal configuration with a bias voltage applied between the source electrode on the TI and the drain (ground) electrode on the BLG, and the current measured between the same two electrodes. In this configuration, in-plane graphene and TI resistance are measured in series with the junction, however they are an order of magnitude smaller than the junction impedance.

Device 1 was measured in a three terminal current bias configuration illustrated in Fig. 1(e). Current is induced between the source electrode on the TI and a drain electrode on the BLG. Voltage is measured between source and  drain electrodes ($V_{SD}$), and also between the source and an additional probe electrode on the graphene ($V_{SP}$). The latter configuration attempts to isolate the junction voltage by removing the in-plane contribution of the graphene \footnote{However, a series in-plane voltage drop on the TI remains as a background to the signal of interest.}. In both devices, gate voltage was applied by a doped Si back-gate through 285 nm of \sio\ dielectric.

Fig.~\ref{Figure_2} compares the differential conductance of the two devices vs. bias voltage. The differential conductance of the high impedance junction in Device 2 traces features which strongly resemble the DOS of the TI surface state, with the Dirac point at $\sim$-230 mV \footnote{A small region of negative differential conductance was observed around the Dirac point, and is as yet unexplained.}. This is similar to the result previously reported in	\cite{Steinberg2015}. For the low impedance junction in Device 1, the differential conductance has no obvious features within the accessible bias range of $|V_{SP}|\leq$ 40 mV (limited due to high currents), other than a suppression around zero bias voltage. 

Device 1 exhibits junction resistance which is at least five times smaller than any previously reported measurements for a mechanically stacked device, along with a qualitatively different \didv\ trace. Variability in the tunneling impedance of graphene-Bi$_2$Se$_3$ heterostructures may arise due to oxidation at the interface, or surface roughness leading to a smaller effective contact area \cite{Steinberg2015}. We raise the conjecture that the strength of coupling between the two materials is also dependent on their relative crystallographic orientations. These may be fortuitously closely aligned in Device 1, accounting for the exceptionally low impedance of the junction.

\begin{figure*}[t]

	\begin{center}
		\includegraphics[width = 170mm]{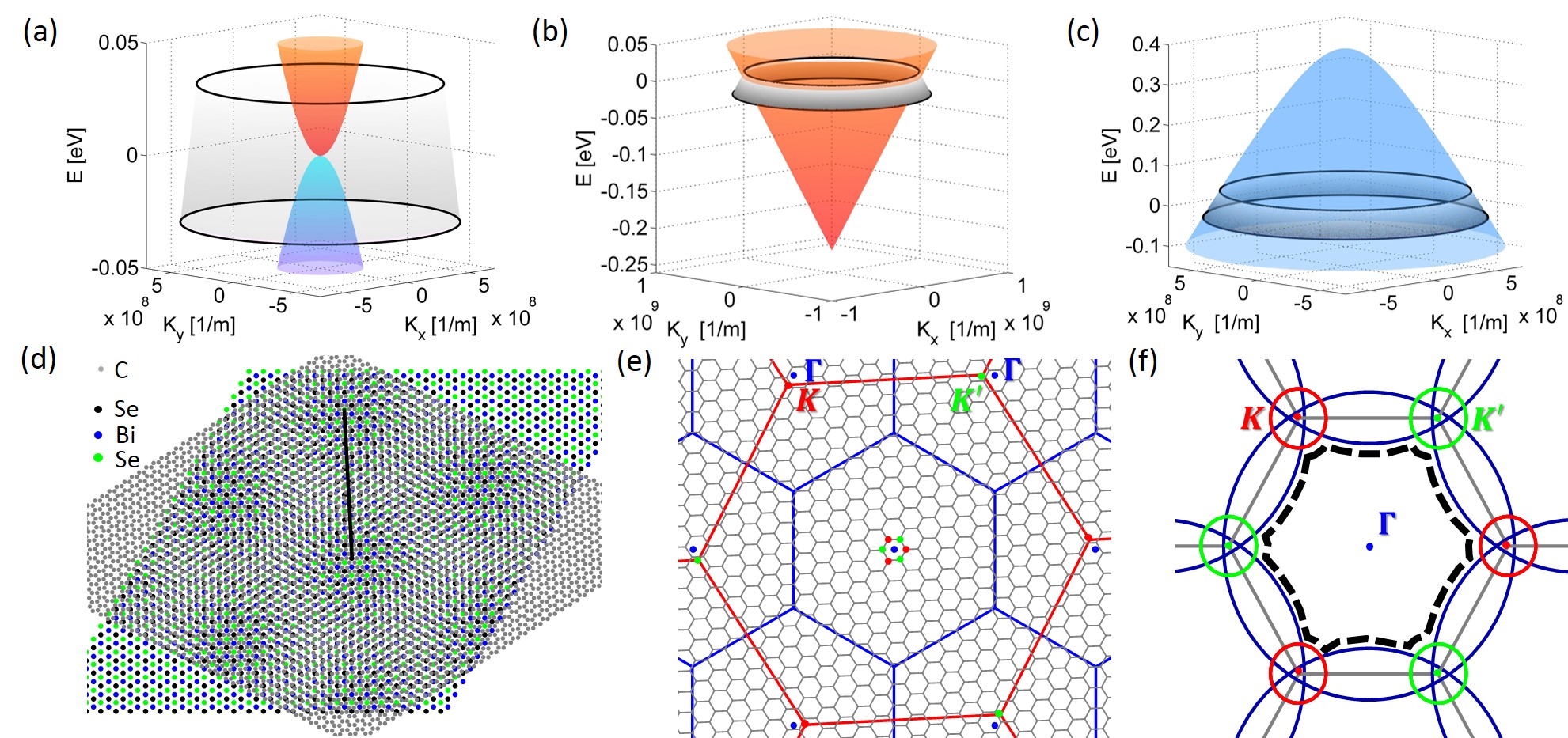}
		\vspace{0cm}
		\caption{(a)-(c) Visualization of measured A$_k$, depicted as circular, in comparison with expected dispersions of (a) pristine BLG, (b) pristine \bise\ , and (c) highly hole-doped BLG. Electron bands are shown in red, hole bands in blue. Gate voltage is roughly translated to an energy scale using typical dispersion parameters discussed in the text. A$_k$ perimeters at the extremal gate voltages of $\pm35$ V are marked in black, and the intermediate dispersion is shaded gray. (d) Real space moir\'e pattern between graphene and \bise\ lattices at a 33$^\circ$ twist angle. The black line shows the SL wavelength of $\approx$ 4 nm. (e)  Momentum space illustration of BLG (red), \bise\ (blue), and SL (gray) BZs at a 33$^\circ$ twist angle, with the BLG K,K' points marked in red and green respectively, and the \bise\ $\Gamma$ point marked in blue (f) Closeup of the central SLBZ. BLG (red/green,holes) and \bise\ (blue, electrons) Fermi pockets repeat with SL periodicity. A possible hole pocket equal in area to the measured $A_k$ exists in the second BZ of the \bise\ dispersion, outlined with a black dotted line.}
		\label{Figure_4}
	\end{center}
	\vspace{-0.5cm}
\end{figure*}

Direct evaluation of the twist angle of buried interface devices such as those reported here is a challenge. Visualization of the moir\'e pattern via STM or AFM measurements, as in references \cite{Li2009a,Dean2013c,Schmidt2014a}, is impossible in this case. We therefore developed an alternative technique based on electron backscatter diffraction (EBSD), which we used to determine the orientation of the \bise\ flake. EBSD, together with the facets on the BLG flake, allow us to estimate the twist angles as 3$^\circ\pm1^\circ$, modulo 30$^\circ$ and  14$^\circ\pm2^\circ$, modulo 30$^\circ$ in Devices 1 and 2 respectively \cite{SI}. The modulo 30$^\circ$ uncertainty is due to armchair/zigzag edge ambiguity of the BLG facets (see Figs.~\ref{Figure_1}(b),(c)). The error given is due to the EBSD measurement. However, there is an additional uncertainty arising from the possibility of ``false faceting'' - straight flake edges which do not exactly follow crystallographic directions. We note that the $\sim$ 33$^\circ$ orientation for Device 1 is near to the $\sqrt{3}$ commensurate angle of 30$^\circ$ and gives rise to a long wavelength moir\'{e} pattern (Fig.~\ref{Figure_4}(d)).

We now report the high magnetic field measurements which constitute the main result of this work. We observe clear quantum oscillations in the zero bias three terminal differential voltage $V_{SP}$, which we explore as a function of gate voltage and magnetic field, revealing two intersecting Landau fans (Fig.~\ref{Figure_3}(a)). These features are apparent only in the junction voltage $V_{SP}$, not in $V_{SD}$ or in the in-plane BLG measurements. One fan originates at a gate voltage of $\sim$ 8 V, and exhibits features at Landau level indices -1,1. This fan is attributed to the disordered BLG on \sio\, which has its neutrality point at $V_g \approx 12 V$ outside the junction area. The second fan, highlighted by the magnetic field derivative of the raw data (Fig.~\ref{Figure_3}(c)), exhibits clear features at Landau level indices N=-22 to -36 and originates at an enormous gate voltage of $\sim$ 294 V. This does not comply with the expected behavior of charge carriers in either pristine BLG or \bise. 

The frequency $F$ of the quantum oscillations in $1/B$ at every gate voltage depends on the area $A_k$ of the Fermi surface in momentum space: $F(V_g) = A_k/{4\pi^2} \times \phi_0$, where $\phi_0 = h/e$ is the magnetic flux quantum. This area decreases with increasing gate voltage, as befits hole carriers. Taking into account the Landau level degeneracy $g$, the charge carrier density $n$ is given by $F = (n/g)h/e$. We may extract the degeneracy from the slopes of the best fit Landau fan (Fig.~\ref{Figure_3}(b)), using the known \sio\ back gate capacitance $C_g$. By this method we obtain an unusual degeneracy of $g$ = 4.95$\pm$0.05. This can be interpreted as a degeneracy of 4 for BLG charge carriers, along with a  $\approx$ 0.8$C_g$ effective gate capacitance due to some of the charge leaking through to the TI \cite{SI}. Conversely, we may assume a degeneracy of 1 for TI charge carriers, along with a $\approx$ 0.2$C_g$ effective gate capacitance due to partial screening by the BLG. Fitting details are given in the supplementary material is. Fig.~\ref{Figure_3}(d)  shows the Fourier spectrum as a function of gate voltage. The oscillation frequency $F$ is translated to hole charge carrier density assuming a degeneracy of 4. A large frequency of $F \approx 200$ T results in the high density of $\approx$ 2$\times$10$^{13}$  cm$^{-2}$ at the Fermi level.  

Seeking the origin of the observed oscillations, we proceed to compare the measured Fermi pockets $A_k$ with typical BLG and TI dispersions (Figs.~\ref{Figure_4}(a),(b)). The effective mass for the BLG dispersion is taken as $m_{eff}$ = 0.04 $\times m_e$ and the graphene Fermi velocity as $v_F$ = 10$^6$ m/s, with the neutrality point at zero energy. The Fermi velocity for the \bise\ surface state dispersion is taken as 5 $\times$ 10$^5$ m/s, with the Dirac point at -0.23 eV (as measured for Device 2 on the same chip). In order to translate the measured $A_k$ vs. $V_g$ roughly to a dispersion relation, we assume that the Fermi pockets are circular, with the radius equal to the Fermi momentum. We then replace the $V_g$ dependence with an inferred energy scale, assigning an energy to each Fermi momentum using the same dispersion parameters described earlier. 

We find that the measured $A_k$ is far larger than the area expected for undoped BLG. It is also larger than expected for the \bise\ surface state (the \bise\ bulk Fermi pocket is smaller still). For comparison, the quantum oscillation frequency $F$ of surface carriers in \bise\ is typically measured to be around 100-120 T for low bulk carrier density samples, and approaches 200 T only for very highly doped samples \cite{Analytis2010a,Taskin2012,Petrushevsky2012}. Moreover, typical \bise\ carriers (both surface and bulk) are electrons, not holes. 

It appears that the interface carriers do not correspond to those of pristine \bise\ or BLG. The simplest plausible explanation for the data is that we are observing behavior of hole-doped BLG in the region of the interface, with charge transfer on the order of 2 $\times$ 10$^{13}$ cm$^{-2}$. This implies that the two fans originate from distinct regions in the sample with different levels of doping. This scenario is illustrated in Fig.~\ref{Figure_4}(c), where the neutrality point of the BLG dispersion is shown shifted upwards by 0.38 eV to fit the measured $A_k$. There are density functional theory calculations showing that monolayer graphene becomes heavily hole-doped in commensurate proximity to few-layer \bise\ \cite{Liu2013a,Spataru2014a}. Experimental observations, however, have not thus far shown an effect of such magnitude. Double-gated transport measurements of BLG-\bise\ mechanically stacked heterostructures do not show an appreciable shift of the neutrality point in the region of the junction \cite{Steinberg2015}. ARPES measurements of similar heterostructures, with BTS grown epitaxially over monolayer CVD graphene, show the monolayer graphene Dirac point shifted by 0.2 eV above the Fermi level \cite{Lee2015a}. This is equivalent to a charge transfer of $\approx$ 3$\times$10$^{12}$ cm$^{-2}$. 

If we assume a nearly commensurate alignment between the two materials, as is possible based on our EBSD measurements, an alternative interpretation could be that we are observing multiple Landau fans due to some form of mini-band transport. It is tempting to interpret the low fill factor fan (the red fan in Fig.~\ref{Figure_3}(b)) as a primary BLG fan and the high fill factor fan (the black fan in Fig.~\ref{Figure_3}(b)) as a secondary BLG fan arising from a high energy secondary neutrality point at the SLBZ boundary, as reported in BLG on hBN \cite{Dean2013c}. However, this interpretation would imply the existence of another secondary fan coming from a negative energy neutrality point, which we do not observe. Also, the fans should in this case terminate at Van Hove singularities (outside of the high magnetic field Hofstadter regime) \cite{Ponomarenko2013a}; whereas we observe two intersecting fans.

Another intriguing interpretation is that we observe a hybrid system, with the low fill factor fan originating from BLG-like carriers, while the high fill factor fan arises from a hole mini-band in the second SLBZ of the \bise\ surface state. Such a system could give rise to the measured degeneracy of 5, with four states from the BLG and an additional state from the TI. To investigate this hypothesis, we super-impose the dispersions of \bise\ and BLG in momentum space. There is a large momentum mismatch between the Dirac point of the \bise\ surface states and the charge neutrality points of the BLG (located at the $\Gamma$ and K,K' points respectively), which should limit hybridization near the Fermi level at large twist angles. 

At near commensurate twist angles however, the momentum-space map drastically changes with the introduction of a small SLBZ, as illustrated in Figs.~\ref{Figure_4}(d)-(f) for an angle of 33$^\circ$ (following references \cite{Bistritzer2011b,Zhang2014b}). At this angle, if the \bise\ Fermi momentum is large enough, the \bise\ dispersion spills out of the first SLBZ at the Fermi level. The second SLBZ ``folds back'' into the first SLBZ to form a hole mini-band with the correct $A_k$. Additionally, the BLG and \bise\ dispersions can intersect in momentum space allowing for hybridization. The feasibility of this interpretation depends strongly on small changes in such parameters as the \bise\ Fermi velocity, Dirac point energy, and the twist angle between the materials. Moreover, the picture in Fig.~\ref{Figure_4}(f) requires the \bise\ Fermi momentum to be around 1.3 times larger than the high end of the range of values reported in \cite{Analytis2010a,Taskin2012,Petrushevsky2012} for pristine \bise, implying some renormalization of the \bise\ surface state dispersion parameters due to hybridization \cite{SI}. 

We conclude that the interface high-density hole carriers do not simply conform to existing theory. Our data encourages consideration of several interesting possibilities at the interface, such as large scale charge transfer, super-lattice effects and hybridization. We believe that these results indicate the existence of a strong coupling regime for BLG-\bise\ devices, and should motivate further theoretical study of twist angle dependent hybrid devices. This as yet unexplained finding should also provide motivation to attempt the further experimental study of controlled twist angle graphene-TI heterostructures. 

The authors are thankful for stimulating discussions with E. Rossi. Electron backscatter diffraction measurements were carried out with the aid of Dr. Z. Barkay at the Wolfson Applied Materials Research Center. Device fabrication and characterization were carried out at the Harvey M. Krueger Family Center for Nanoscience and Nanotechnology. Work was supported by the Israel Science Foundation Grant No. 2103/15, Marie Curie CIG Grant No. PCIG12-GA-2012-333620 and ERC-2014-STG Grant No. 637298.


\begin{thebibliography}{10}
	
	\bibitem{Zhang2014b}
	J.~Zhang, C.~Triola, E.~Rossi:
	\newblock \emph{Phys. Rev. Lett.} \textbf{112}  096802 (2014)
	
	\bibitem{Jin2013a}
	K.~H. Jin, S.~H. Jhi:
	\newblock \emph{Phys. Rev. B}
	\textbf{87}  075442 (2013)
	
	\bibitem{Cao2016}
	W.~Cao \emph{et al.}:
	\newblock \emph{2D Materials} \textbf{3}  034006 (2016) 
	
	\bibitem{Popov2014}
	I.~Popov, M.~Mantega, A.~Narayan, S.~Sanvito:
	\newblock \emph{Phys. Rev. B}
	\textbf{90}  035418 (2014) 
	
	\bibitem{Kou2013}
	L.~Kou \emph{et al.}:
	\newblock \emph{Nano Lett.} \textbf{13}  6251 (2013) 
	
	\bibitem{Liu2013a}
	W.~Liu \emph{et al.}:
	\newblock \emph{Phys. Rev. B}
	\textbf{87}  205315 (2013) 
	
	\bibitem{Spataru2014a}
	C.~D. Spataru, F.~L{\'{e}}onard:
	\newblock \emph{Phys. Rev. B} \textbf{90}  085115 (2014) 
	
	\bibitem{Li2012}
	H.~Li \emph{et al.}:
	\newblock \emph{Appl. Phys. Lett.} \textbf{101}  243102 (2012) 
	
	\bibitem{Steinberg2015}
	H.~Steinberg \emph{et al.}:
	\newblock \emph{Phys. Rev. B} \textbf{92}  241409(R) (2015) 
	
	\bibitem{Zhang2016a}
	L.~Zhang, Y.~Yan, H.~C. Wu, D.~Yu, Z.~M. Liao:
	\newblock \emph{ACS Nano} \textbf{10}  3816 (2016) 
	
	\bibitem{Vaklinova2016a}
	K.~Vaklinova, A.~Hoyer, M.~Burghard, K.~Kern:
	\newblock \emph{Nano Lett.} \textbf{16}  2595 (2016) 
	
	\bibitem{tian2016electrical}
	J.~Tian, T.~F. Chung, I.~Miotkowski, Y.~P. Chen:
	\newblock \emph{arXiv:1607.02651}  (2016)
	
	\bibitem{Park2008}
	C.~H. Park, L.~Yang, Y.~W. Son, M.~L. Cohen, S.~G. Louie:
	\newblock \emph{Nature Phys.} \textbf{4}  213 (2008) 
	
	\bibitem{Yankowitz2012a}
	M.~Yankowitz \emph{et al.}:
	\newblock \emph{Nature Phys.} \textbf{8}  382 (2012) 
	
	\bibitem{Hunt2013b}
	B.~Hunt \emph{et al.}:
	\newblock \emph{Science} \textbf{340}  1427 (2013) 
	
	\bibitem{Ponomarenko2013a}
	L.~A. Ponomarenko \emph{et al.}:
	\newblock \emph{Nature} \textbf{497}  594 (2013) 
	
	\bibitem{LopesDosSantos2007a}
	J.~M.~B. Lopes Dos Santos, N.~M.~R. Peres, A.~H. Castro Neto:
	\newblock \emph{Phys. Rev. Lett.} \textbf{99}  256802 (2007) 
	
	\bibitem{Mele2010}
	E.~J. Mele:
	\newblock \emph{Phys. Rev. B}
	\textbf{81}  161405(R) (2010) 
	
	\bibitem{Bistritzer2011c}
	R.~Bistritzer and A.~H. MacDonald:
	\newblock \emph{Proc. National Acad. Sciences United
		States Am.} \textbf{108}  12233 (2011) 
	
	\bibitem{Ohta2012}
	T.~Ohta \emph{et al.}:
	\newblock \emph{Phys. Rev. Lett.} \textbf{109}  186807 (2012) 
	
	\bibitem{Schmidt2014a}
	H.~Schmidt, J.~C. Rode, D.~Smirnov, R.~J. Haug:
	\newblock \emph{Nature Commun.} \textbf{5}  5742 (2014) 
	
	\bibitem{Cao2016a}
	Y.~Cao \emph{et al.}: 
	\newblock \emph{Phys. Rev. Lett.} \textbf{117}  116804 (2016) 
	
	\bibitem{Kim2016}
	Y.~Kim \emph{et al.}:
	\newblock \emph{Nano Lett.} \textbf{16}  5053 (2016) 
	
	\bibitem{Song2010a}
	C.~L. Song \emph{et al.}:
	\newblock \emph{Appl. Phys. Lett.} \textbf{97}  143118 (2010) 
	
	\bibitem{Li2009a}
	G.~Li \emph{et al.}:
	\newblock \emph{Nature Phys.} \textbf{6}  109 (2010) 
	
	\bibitem{Dean2013c}
	C.~R. Dean \emph{et al.}:
	\newblock \emph{Nature} \textbf{497}  598 (2013) 
		
	\bibitem{Analytis2010a}
	J.~G. Analytis \emph{et al.}:
	\newblock \emph{Phys. Rev. B}
	\textbf{81}  205407 (2010) 
	
	\bibitem{Taskin2012}
	A.~A. Taskin, S.~Sasaki, K.~Segawa, Y.~Ando:
	\newblock \emph{Phys. Rev. Lett.} \textbf{109}  066803 (2012) 
	
	\bibitem{Petrushevsky2012}
	M.~Petrushevsky \emph{et al.}:
	\newblock \emph{Phys. Rev. B}
	\textbf{86}  045131 (2012) 
	
	\bibitem{Lee2015a}
	P.~Lee \emph{et al.}:
	\newblock \emph{ACS Nano} \textbf{5} 10861 (2015) 
	
	\bibitem{Bistritzer2011b}
	R.~Bistritzer and A.~H. MacDonald:
	\newblock \emph{Phys. Rev. B}
	\textbf{84}  035440 (2011) 
	
	\bibitem{SI}
	\bibinfo{note}{See supplementary material}.
	
\end{thebibliography}
\end{document}